\documentclass[11pt]{article}

\usepackage{amssymb,amsfonts,amsmath}

\usepackage{pstricks,pst-node,pst-text,pst-3d}

\newtheorem{theorem}{Theorem}[section]
\newtheorem{lemma}[theorem]{Lemma}
\newtheorem{proposition}[theorem]{Proposition}
\newtheorem{example}[theorem]{Example}
\newtheorem{corollary}[theorem]{Corollary}
\newtheorem{remark}[theorem]{Remark}

\newcommand{\bp}{\noindent {\em Proof. }}
\newcommand{\ep}{\hfill $\Box$ \par\medskip}

\def\calA{{\mathcal A}}
\def\calB{{\mathcal B}}

\def\calF{{\mathcal F}}
\def\calG{{\mathcal G}}

\def\calN{{\mathcal N}}
\def\calO{{\mathcal O}}

\def\bfA{{\mathbf A}}
\def\bfB{{\mathbf B}}

\def\bfI{{\mathbf I}}

\def\bf1{{\mathbf 1}}
\def\bf2{{\mathbf 2}}

\def\ra{\rightarrow}

\def\es{\emptyset}
\def\se{\subseteq}

\def\ve{\varepsilon}

\def\vp{\varphi}

\def\Eq{\mathrm{Eq}}
\def\lg{\mathrm{lg}}

\def\faA{\mathcal{A} = (A,X,\delta)}
\def\ndA{\mathcal{N} = (A,X)}

\def\fzA{\mathcal{F} = (A,X,f)}

\def\hF{\widehat{\calF}}

\def\FDir{\mathbf{FDir}}
\def\CFDir{\mathbf{CFDir}}

\begin{document}

\title{Directable Fuzzy and Nondeterministic Automata}

\author{Magnus Steinby\\
Department of Mathematics and Statistics\\
University of Turku\\
20014 Turku, Finland}
\date{September 2017}

\maketitle

\begin{abstract}
We study three notions of directability of fuzzy automata akin to the D1-, D2- and D3-directability of nondeterministic automata. Thus an input word $w$ of a fuzzy automaton is D1-directing if a fixed single state is reachable by $w$ from all states, D2-directing if exactly the same states are reachable by $w$ from every state, and D3-directing if there is a state reachable by $w$ from every state. We study the various sets of directing words of fuzzy automata, prove that the directability properties are decidable, and show how such results can be
deduced from the theory of directable nondeterministic automata. Moreover, we establish the closure properties of the different classes of directable fuzzy automata under the class operations of forming subautomata, homomorphic images and finite direct products.
\end{abstract}

\smallskip


\section{Introduction}\label{Introduction}

An input word $w$ of a finite automaton $\calA$ is said to be \emph{directing}, or \emph{synchronizing}, if it takes $\calA$ from every state to a fixed state, and $\calA$ is \emph{directable} if it has directing words. The topic has been extensively studied ever since these notions were introduced in 1964 by \v{C}ern\'{y} \cite{Cer64}. Directability has also been defined for other kinds of automata such as probabilistic automata \cite{Kfo70} and weighted automata \cite{DoJuLaMaSh13,Iva14}. In \cite{ImSt99} Imreh and Steinby identify three types of directable nondeterministic finite automata (NFAs), and in \cite{KaRa15} \text{Karthikeyan} and \text{Rajasekar} define  directing words for fuzzy automata. Speranskii \cite{Spe15} also considers a kind of generalized directing words of fuzzy automata as well as some diagnostic experiments with fuzzy machines with outputs. Here we introduce and study three types of directing words of fuzzy automata that correspond to the D1-, D2- and D3-directing words NFAs considered in \cite{ImSt99}.

The paper is organized as follows. In Section 2 we  introduce some terminology, notation and the automata to be considered. In Section 3 we associate with each fuzzy automaton an NFA that faithfully reflects some important properties of the original automaton. In the next section we define D1-, D2- and D3-directing words for fuzzy automata. The D3-directing words of a fuzzy automaton are the directing words defined in \cite{KaRa15}. For each $i = 1,2,3$, the D$i$-directing words of a fuzzy automaton are the same as the D$i$-directing words of the associated nondeterministic automaton, and hence many facts about fuzzy automata and their directing words follow directly from the existing theory of directable nondeterministic automata. In Section 5 it is shown that the D3-directability of a complete fuzzy automaton\footnote{A fuzzy automaton is said to be complete if for every state-input pair there is at least one nonzero transition.} can be tested by an algorithm that is essentially the same as the one presented in \cite{ImSt95} for testing the directability of an ordinary finite automaton. In Section 6 we establish the closure properties of the classes of D1-, D2- and D3-directable fuzzy automata and complete fuzzy automata under the operations of forming subautomata, homomorphic images and finite direct products. In Section 7 we note a couple of possible extensions of this work.

We assume that the reader is familiar with the basic theory of
finite automata and regular languages (cf.  \cite{HoMoUl07},  \cite{Koz97} or \cite{Sak09}, for example).  For the general theory
of fuzzy automata and fuzzy languages, the reader may consult
\cite{MoMa02} or \cite{Rah09}.

\section{Some basic notions}\label{Basic}

Sometimes we use the notation $A := B$ to emphasize that $A$ is defined to be equal to $B$. The cardinality of a set $A$ is denoted by $|A|$.
For any mapping $\vp : A \ra B$, the value $\vp(a)$ of an element $a\in A$ is also written as $a\vp$, and for any $H \se A$, the set $\vp(H) := \{a\vp \mid a\in H\}$ may be denoted by $H\vp$. 
For a relation $\theta \se A\times B$, we may express $(a,b)\in \theta$ also by writing $a\, \theta \, b$. For any $a\in A$ and $H \se A$, $a\theta := \{b\in B \mid a\, \theta \,b\}$ and $H\theta := \bigcup\{a\theta \mid a \in H\}$.
The diagonal relation $\{(a,a) \mid a\in A\}$ is denoted by $\Delta_A$. The set of equivalence relations on a set $A$ is denoted by $\Eq(A)$. If $\theta \in \Eq(A)$ is known from the context, we may denote the $\theta$-class $a\theta$ of an element $a\in A$ by $[a]$.

In what follows, $X$ is always a finite nonempty alphabet. The set of all (finite) words over $X$ is denoted by $X^*$ and the empty word by $\ve$. The length of a word $w \in X^*$ is denoted by $\lg(w)$. Subsets of $X^*$ are called \emph{languages}.

We consider automata without initial states and outputs because directing words are defined without any reference to these.
Thus a \emph{deterministic finite automaton}, a \emph{DFA} for short, $\faA$ consists of a finite nonempty set $A$ of \emph{states}, the \emph{input alphabet} $X$, and a \emph{transition function} $\delta : A\times X \ra A$. As usual, $\delta$ is extended to a mapping $\delta^* : A\times X^* \ra A$ by setting, for every $a\in A$, $\delta^*(a,\ve) = a$ and $\delta^*(a,w) = \delta(\delta^*(a,v),x)$ for $w = vx$ with $v\in X^*$ and $x\in X$. As there is no danger of confusion, we  write $\delta(a,w)$ instead of $\delta^*(a,w)$.

A word $w\in X^*$ is a \emph{directing word} of a DFA $\faA$ if there is a state $c\in A$ such that $\delta(a,w) = c$ for every $a\in A$. The set of directing words of $\calA$ is denoted by $DW(\calA)$, and $\calA$ is \emph{directable} if $DW(\calA) \neq \es$. There is an extensive literature on directable automata. We mention just the surveys \cite{BoImCiPe99} and \cite{Vol08}, and the paper \cite{ImSt95} to which we shall refer several times.

A \emph{nondeterministic finite automaton (NFA)} $\ndA$ is a system in which $A$ is a finite nonempty set of \emph{state}s, $X$ is the \emph{input alphabet}, and each input letter $x\in X$ is realized as a \emph{transition relation} $x^\calN \se A \times A$. For any $a\in A$ and $x\in X$, $ax^\calN
:= \{ b\in A \mid (a,b)\in x^\calN\} $ is the set
of states which $\calN$ may enter
from state $a$ by reading the input letter $x$.
For any $H\subseteq A$ and $w\in X^*$, the set $Hw^\calN$
is obtained inductively by
\begin{itemize}\vspace{-3pt}
  \item[{\rm (1)}] $H\ve ^\calN := H$, $Hx^\calN := \bigcup\{ax^\calN \mid a \in H\}$ for $x\in X$, and \vspace{-3pt}
  \item[{\rm (2)}] $Hw^\calN := (Hv^\calN)x^\calN$ for $w=vx$ with $v\in X^*$ and $x\in X$. \vspace{-3pt}
\end{itemize}
For any $a\in A$, let $aw^\calN := \{ a\} w^\calN$.
We call $\calN$ a {\it complete NFA}, if
$ax^\calN\not =\emptyset $ for all $a\in A$ and $x\in X$.
We defined NFAs as relational structures, but we may still view a DFA $\faA$ as a special NFA in which each relation $x^\calA := \{(a,\delta(a,x)) \mid a\in A\}$ ($x\in X$) is a function.

A \emph{fuzzy automaton} $\fzA$ consists of   a finite nonempty set $A$ of \emph{states}, the \emph{input alphabet} $X$, and  a fuzzy \emph{transition function} \vspace{-3pt}
$$
f : A\times X \times A \ra [0,1], \vspace{-3pt}
$$
where $[0,1]$ is the closed real interval between $0$ and $1$. The transition function is extended to  $f^* : A\times X^* \times A \ra [0,1]$ by
\begin{itemize} \vspace{-3pt}
  \item[(1)] $f^*(a,\ve,a) = 1$, and $f^*(a,\ve,b) = 0$ if $b\neq a$, and  \vspace{-3pt}
  \item[(2)] $f^*(a,w,b) = \max\{\min\{f^*(a,v,c),f(c,x,b)\}\mid c\in A\}$
  for $w = vx$ with $v\in X^*$ and $x\in X$,  \vspace{-3pt}
\end{itemize}
for all $a,b\in A$. We may write just $f(a,w,b)$ for $f^*(a,w,b)$.

For any $a\in A$ and $w\in X^*$, let $\calF(a,w) := \{b\in A \mid f(a,w,b) > 0\}$ be the set of all states that $\calF$ reaches from  $a$ by the input word $w$. For any $H\se A$ and $w\in X^*$, let $\calF(H,w) := \bigcup\{\calF(a,w) \mid a\in H\}$. The fuzzy automaton $\calF$ is \emph{complete} if $\calF(a,x) \neq \es$ for all $a\in A$ and $x\in X$. It is clear that if $\calF$ is complete, then $\calF(a,w) \neq \es$ for all $a\in A$ and $w\in X^*$.

The following obvious lemma has an easy proof by induction on $\lg(v)$.

\begin{lemma}\label{le:F(F(H,uv)) = F(F(H.u),v)} $\calF(H,uv) = \calF(\calF(H,u),v)$ for all $H \se A$ and $u,v\in X^*$. In particular, $\calF(a,uv) = \calF(\calF(a,u),v)$ for any $a\in A$ and $u,v\in X^*$.
\end{lemma}

\section{From fuzzy to nondeterministic automata}\label{se:Fuzzy automata and NFA}

It is obvious that the directability of fuzzy automata could be defined in several meaningful ways. The definition suggested by \text{Karthikeyan} and \text{Rajasekar} \cite{KaRa15} may be stated as follows: $w\in X^*$ is a \emph{directing word} of a fuzzy automaton $\fzA$ if there is a state $c\in A$ such that $f(a,w,c) > 0$ for every $a\in A$, and   $\calF$ is said to be \emph{directable} if it has a directing word. In \cite{KaRa15} some ideas and results appearing in \cite{ImSt95} are presented in a `fuzzified' form using these notions. The discussion of these and related matters is simplified by the following reduction to nondeterministic automata.

Consider a fuzzy automaton $\fzA$ and a word $w = x_1\ldots x_k$ $(x_1,\ldots,x_k\in X)$.
Clearly, $w$ is directing for $\calF$ if and only if there is a state $c\in A$ such that for every $a\in A$ there is a chain of states $a = a_0,a_1,\ldots,a_k = c$ such that $f(a_{i-1},x_i,a_i) > 0$ for every $i = 1,\ldots,k$. The actual values of the numbers $f(a_{i-1},x_i,a_i)$ do not matter as long as they are positive. Hence, we could assign every positive transition the value 1 without changing the set of directing words. This means that when we are interested just in directing words, $\calF$ may be replaced by the \emph{associated NFA} $\hF =(A,X)$, where $ax^{\hF} = \{b\in A \mid f(a,x,b) > 0\}$ for all $x\in X$ and $a\in A$.

\begin{example}\label{ex:Fuzzy to NFA}{\rm Let $\calF = (\{a,b,c\},\{x,y\},f)$ be the fuzzy automaton, where the non-zero transitions are $f(a,x,b) = 0.3$, $f(b,x,c) = 0.4$, $f(c,x,b) = 0.2$, $f(c,x,c) = 0.6$, $f(b,y,b) = 0.5$ and $f(b,y,c) = 0.1$.
Then the transition relations of $\hF$ are given by $ax^{\hF} = \{b\}$, $bx^{\hF} = \{c\}$, $cx^{\hF} = \{b,c\}$, $ay^{\hF} = \es$, $by^{\hF} = \{b,c\}$ and $cy^{\hF} = \es$.
}
\end{example}

It is easy to verify the following lemma by induction on $\lg(w)$.

\begin{lemma}\label{le:fzA to ndA} Let $\fzA$ be a fuzzy automaton. Then $aw^{\hF} = \calF(a,w)$  for all $a\in A$ and $w\in X^*$. In other words, for all $w\in X^*$ and  $a,b\in A$, $b \in aw^{\hF}$ if and only if $f(a,w,b) > 0$.
\end{lemma}

Let us also note the following obvious fact.

\begin{corollary}\label{co:Completeness of automata} A fuzzy automaton $\calF$ is complete if and only if the associated NFA $\hF$ is complete.
\end{corollary}

\section{Directing words and directable automata}\label{se:DirWords}

In \cite{ImSt99} three kinds of directing words of an NFA $\ndA$ were identified. A word $w\in X^*$ was said to be
\begin{itemize} \vspace{-3pt}
  \item[{\rm (D1)}] \emph{D1-directing} if $(\exists c\in A)(\forall a\in A)(aw^\calN= \{ c\} )$, \vspace{-3pt}
  \item[{\rm (D2)}] \emph{D2-directing} if $(\forall a,b\in A)(aw^\calN = bw^\calN)$, and \vspace{-3pt}
  \item[{\rm (D3)}] \emph{D3-directing} if $(\exists c\in A)(\forall a\in A)(c\in aw^\calN)$. \vspace{-3pt}
\end{itemize}
For each $i=1,2,3$, the set of  D$i$-directing words of $\calN$ is denoted by $D_i(\calN)$, and $\calN$ is called D$i$-\emph{directable} if $D_i(\calN)\neq \es$. Note that for a DFA, all three conditions (D$i$) yield the usual directing words.

For a fuzzy automaton $\fzA$, let us call a word $w\in X^*$
\begin{itemize} \vspace{-3pt}
  \item[{\rm (D1')}] \emph{D1-directing} if $(\exists c\in A)(\forall a\in A)(\calF(a,w) = \{c\} )$, \vspace{-3pt}
  \item[{\rm (D2')}] \emph{D2-directing} if $(\forall a,b\in A)(\calF(a,w) = \calF(b,w))$, and\vspace{-3pt}
  \item[{\rm (D3')}] \emph{D3-directing} if $(\exists c\in A)(\forall a\in A)(c\in \calF(a,w))$. \vspace{-3pt}
\end{itemize}
For each $i=1,2,3$, the set of  D$i$-directing words of $\calF$ is denoted by $D_i(\calF)$, and $\calF$ is called D$i$-\emph{directable} if $D_i(\calF)\neq \es$. The classes of D$i$-directable fuzzy automata and complete D$i$-directable fuzzy automata are denoted by $\FDir(i)$ and $\CFDir(i)$, respectively.

The D3-directing words of a fuzzy automaton $\calF$ are exactly the directing words of $\calF$ as defined  in \cite{KaRa15}. In \cite{Spe15} Speranskii calls  a word $w\in X^*$ a ``generalized synchronizing word'' of $\calF$ if $\calF(A,w)$ is a proper subset of $A$. If $\calF$ is complete, such a  $w$ is a D1-directing word in our sense in the special case $|\calF(A,w)| = 1$.

A pairwise comparison of the conditions (Di) and (Di') ($i=1,2,3$)  yields by Lemma \ref{le:fzA to ndA} the following facts which imply that all results concerning directing words of NFAs apply to fuzzy automata, too.

\begin{proposition}\label{pr:Di(fzA) = Di(ndA)} For any fuzzy automaton $\calF$ and each $i = 1,2,3$,
$D_i(\calF) = D_i(\hF)$, and hence
$\calF$ is Di-directable if and only if $\hF$ is Di-directable.
\end{proposition}

Clearly, (D1) is the strongest of the three types of directability: exactly one state is reachable by a D1-directing input word and this state does not depend on the starting state. On the other hand, $w\in D_2(\calF)$ means just that the set of states reachable by input $w$ is independent of the starting state, and for $w\in D_3(\calF)$, there is a state which is  reachable by $w$ from every state.
The following relations between the sets $D_1(\calF)$, $D_2(\calF)$, and $D_3(\calF)$, for any fuzzy automaton $\calF$, hold by Remark 3.2 of \cite{ImSt99} and Proposition \ref{pr:Di(fzA) = Di(ndA)}.

\begin{remark}\label{re:D1F,D2F and D3F} $D_1(\calF) \se D_2(\calF)\cap D_3(\calF)$ for any fuzzy automaton $\calF$, and if $\calF$ is complete, then $D_1(\calF) \se D_2(\calF) \se D_3(\calF)$. Moreover, any one of the inclusions may be proper.
\end{remark}

From \cite{ImSt99} we get by Proposition \ref{pr:Di(fzA) = Di(ndA)} the Hasse diagram of Figure 1 which shows the inclusion relations between the various classes of directable fuzzy automata. Proposition \ref{pr:Di(fzA) = Di(ndA)} means also that the many results (cf. \cite{Bur76,GaIvNG09,GoHeKoRy82,ImSt99}) concerning bounds of the lengths of the shortest D$i$-directing words of an $n$-state D$i$-directable NFA apply to fuzzy automata, too.

\begin{figure}[t!]
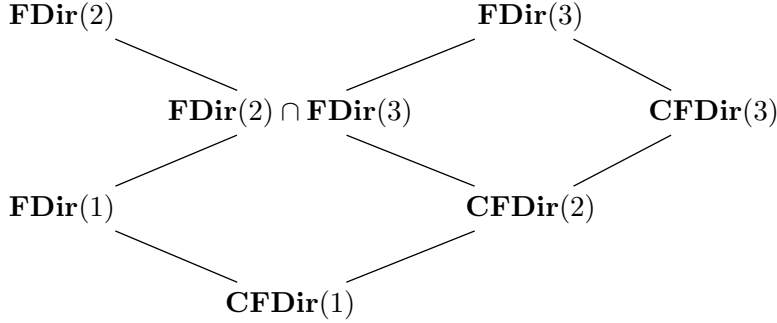


\begin{center}
$\begin{psmatrix}[colsep=7mm,rowsep=8mm]
\rnode{2}{\FDir(2)} & & \rnode{3}{\FDir(3)}\\
& \rnode{23}{\FDir(2)\cap\FDir(3)} &  & \rnode{c3}{\CFDir(3)}\\
\rnode{1}{\FDir(1)} & & \rnode{c2}{\CFDir(2)}\\
& \rnode{c1}{\CFDir(1)}
\end{psmatrix}
$
\psset{arrows=-, nodesep=3pt,linestyle=solid,linewidth= 0.5pt, arrowscale=1.5}
\everypsbox{\scriptstyle}
\ncline{2}{23}
\ncline{3}{23}
\ncline{3}{c3}
\ncline{23}{1}
\ncline{23}{c2}
\ncline{c3}{c2}
\ncline{1}{c1}
\ncline{c2}{c1}
\end{center}

\noindent

\caption{Inclusion diagram}\label{diagr1}
\end{figure}

In \cite{ImSt95} it was noted that $X^*DW(\calA)X^* = DW(\calA)$ for any DFA $\calA$, but the corresponding fact does not always hold for the D3-directing words of a fuzzy automaton $\calF$ as claimed in \cite{KaRa15}. In fact, not even $X^*D_3(\calF) = D_3(\calF)$ or $D_3(\calF)X^* = D_3(\calF)$ holds generally. For example, the fuzzy automaton $\calF$ of Example \ref{ex:Fuzzy to NFA} has the D3-directing word $xx$ while neither $yxx$ nor $xxy$ is in $D_3(\calF)$. Remark 3.3 of \cite{ImSt99} yields the following rather evident facts.

\begin{remark}\label{re:X*DW(F)X* = DW(F)} $D_2(\calF)X^* = D_2(\calF)$ for any fuzzy automaton $\calF$. If $\calF$ is complete, then $X^*D_1(\calF) = D_1(\calF)$, $X^*D_2(\calF)X^* = D_2(\calF)$ and $X^*D_3(\calF)X^* = D_3(\calF)$.
\end{remark}

Recall that a \emph{finite deterministic recognizer} $\bfA = (A,X,\delta,a_0,F)$ is a DFA $(A,X,\delta)$ supplemented with an \emph{initial state} $a_0\in A$ and a set $F\se A$ of \emph{final states}. It \emph{recognizes} the language $L(\bfA) := \{w\in X^* \mid \delta(a_0,w) \in F\}$, and a language $L\se X^*$ is  \emph{regular} if $L = L(\bfA)$ for such a recognizer $\bfA$.

It is rather obvious that  $D_1(\calF)$, $D_2(\calF)$ and $D_3(\calF)$ are regular languages for every fuzzy automaton $\fzA$. However, the simple power set recognizer construction used  in \cite{KaRa15} does not suffice.  In this recognizer the state set is the power set $\wp(A)$ and the fuzzy transition function $g$ is defined by
$g(B,x,C) \,=\, \min\{f(b,x,c) \mid b\in B, c\in C\}$ $(B,C\se A,\, x\in X)$.
Since $A$ is the only initial state, final states are reached in exceptional cases only, and the problem persists even if we consider complete fuzzy automata only. For example, if $A = \{a,b,c\}$, $X = \{x\}$ and the nonzero transitions are given by $f(a,x,b) = f(b,x,c) = f(c,x,c) = 1$, then  $g(A,xx,B) = 0$ for every nonempty $B\in \wp(A)$ although $xx \in D_3(\calF)$. No modification of the transition function would help either because it does not suffice to know what states are reachable by a given input, but we have to know also what states are reachable from each state.

The regularity of the sets $D_i(\calF)$ follows directly from Proposition 3.4 of \cite{ImSt99}, but let us adapt the construction used there for fuzzy automata.

\begin{proposition}\label{pr:DW(A) regular} For any (effectively given) fuzzy automaton $\calF$, the languages $D_1(\calF)$, $D_2(\calF)$ and $D_3(\calF)$ are (effectively) regular.
\end{proposition}

\bp If $\fzA$ with $A = \{a_1,\ldots,a_n\}$, let $\calB = (B,X,\delta)$ be the DFA with $B=\{\{\calF(a_1,u),\dots ,\calF(a_n,u)\} \mid u\in X^*\}$ and $\delta$ defined by \vspace{-3pt}
$$
\delta(\{C_1,\dots,C_k\},x)= \{\calF(C_1,x),\dots,\calF(C_k,x)\}\vspace{-3pt}
$$
for all $\{C_1,\dots,C_k\} \in B$ and $x\in X$.
For $b_0=\{\{ a_1\},\dots,\{a_n\}\}$, we get
$\delta(b_0,u)=\{\calF(a_1,u),\dots,\calF(a_n,u)\}$ for every $u\in X^*$. It is easy to see that, for each $i =1,2,3$, $L(\bfB_i)= D_i(\calF)$
for $\bfB_i =(B,X,\delta,b_0,F_i)$, when we choose $F_1 = \{\{\{c\}\} \mid c\in A\}\cap B$, $F_2 = \{\{C\} \mid C \se A\}\cap B$, and
$F_3 =\{\{C_1,\dots,C_k\}\in B \mid C_1\cap\dots\cap C_k\neq \es \}$.
The construction is clearly effective if $\calF$ is effectively given.\ep


\section{Testing for directability}\label{se:Testing}

Two states $a$ and $b$ of a DFA $\faA$ are \emph{merged} by a word $w\in X^*$ if $\delta(a,w) = \delta(b,w)$. The algorithm for testing the directability of a DFA $\calA$ presented in \cite{ImSt95} is based on the obvious fact that  $\calA$ is directable if and only if every pair of states of $\calA$ can be merged.

By the definition suggested in \cite{KaRa15}, two states $a$ and $b$ of a fuzzy automaton $\fzA$ are \emph{merged} by a word $w\in X^*$ if $f(a,w,c) >0$ and $f(b,w,c) >0$ for some $c\in A$. Theorem 4.1 of \cite{KaRa15} claims that a fuzzy automaton is D3-directable if and only if every pair of states can be merged. The condition is trivially necessary since any directing word merges all pairs of states, but the converse may fail if the automaton is incomplete and then the directability testing algorithm of \cite{KaRa15}, obtained from that of \cite{ImSt95} by replacing the Boolean matrix by a fuzzy matrix, may incorrectly proclaim the fuzzy automaton to be D3-directable.

\begin{example}\label{ex:Dir and merging in fzA}{\rm Let $\calF = (\{a,b,c\},\{x,y,z\},f)$ be a fuzzy automaton in which $f(a,x,a)$, $f(b,x,a)$, $f(b,y,b)$, $f(c,y,b)$, $f(a,z,c)$ and $f(c,z,c)$ are the transitions with non-zero weights.
As $a$ and $b$ are merged by $x$, $b$ and $c$ by $y$, and $a$ and $c$ by $z$, the algorithm of \cite{KaRa15} would claim $\calF$ to be D3-directable. However, $D_3(\calF) = \es$ since $f(a,yw,d) = f(b,zw,d) = f(c,xw,d) = 0$ for all $w\in X^*$ and $d\in A$.
}
\end{example}

The problem is not revealed by the example in \cite{KaRa15} because the fuzzy automaton considered is, in fact, a DFA with positive weights attached to all transitions, and hence its directability could be decided by the original algorithm of \cite{ImSt95} ignoring the weights. We shall show that this algorithm can be used, almost without any change, for testing a complete fuzzy automaton for D3-directability.

In \cite{ImSt99} a word $w\in X^*$ was said to \emph{D3-merge} two states $a,b\in A$ of an NFA $\ndA$ if $aw^\calN\cap bw^\calN \neq \es$. Let us say that a word $w\in X^*$ \emph{D3-merges} two states $a,b\in A$ of a fuzzy automaton $\fzA$ if $\calF(a,w)\cap\calF(b,w) \neq \es$. Obviously, a word merges, in the above sense of \cite{KaRa15}, two states if and only if it D3-merges them.

\begin{lemma}\label{le:Dir and merge in fzA} A word D3-merges two states of a fuzzy automaton $\calF$ if and only if it D3-merges them in the associated NFA $\hF$, and hence a complete fuzzy automaton is D3-directable if and only if every pair of its states has a D3-merging word.
\end{lemma}

\bp For any input word $w$ and any states $a$ and $b$, $\calF(a,w) = aw^{\hF}$ and $\calF(b,w) = bw^{\hF}$ by Lemma \ref{le:fzA to ndA}, and hence $w$ D3-merges $a$ and $b$ in $\calF$ exactly in case it D3-merges them in $\hF$. The rest of the lemma follows now from Lemma 5.4 of \cite{ImSt99} which expresses the corresponding fact for NFAs. \ep

For any $k\geq 0$, two states $a,b\in A$ of a fuzzy automaton $\fzA$ are said to be \emph{D3 $k$-mergeable} if they are D3-merged by a word of length $\leq k$. Assuming  that $\calF$ is known from the context, we  denote the D3 $k$-mergeability relation of $\calF$ by  $\mu(k)$, and let $\mu := \bigcup_{k\geq 0}\mu(k)$.

The following proposition, which justifies the algorithm to be presented, corresponds exactly to  Proposition 4.1 of \cite{ImSt95}.

\begin{proposition}\label{pr:Mu-relations} Let $\fzA$ be a complete fuzzy automaton.
\begin{itemize} \vspace{-3pt}
  \item[{\rm (a)}] $\calF$ is D3-directable if and only if $\mu = A\times A$. \vspace{-3pt}
  \item[{\rm (b)}] The relations $\mu(k)$ and $\mu$ are reflexive and symmetric. \vspace{-3pt}
  \item[{\rm (c)}] $\mu(0) = \Delta_A$, and $\mu(k) = \mu(k-1) \cup\\  \{(a,b) \mid (\exists x\in X)[(\calF(a,x) \times \calF(b,x))\cap \mu(k-1) \neq \es]\}$ for $k\geq 1$. \vspace{-3pt}
  \item[{\rm (d)}] If $\mu(k) = \mu(k-1)$ for some $k\geq 1$, then $\mu(k) = \mu$. \vspace{-3pt}
  \item[{\rm (e)}] $\Delta_A = \mu(0) \subset \mu(1) \subset \ldots \subset \mu(k) = \mu$ for some $k \leq {n \choose 2}$, where $n = |A|$. \vspace{-3pt}
\end{itemize}
\end{proposition}

Let $\fzA$ be an $n$-state complete fuzzy automaton with  $A = \{1,\ldots,n\}$. The algorithm implicitly computes the relations $\mu(0), \mu(1), \ldots$ until $\mu(k-1) = \mu(k)$ for some $k\geq 1$. It employs a Boolean $n\times n$-matrix $M$ and a list $NewPair$ of pairs of states. That $M[i,j] = 1$ means that states $i$ and $j$ are known to be D3-mergeable. It suffices to consider the entries $M[i,j]$ with $1\leq i < j \leq n$. A pair $(i,j)$ appears on the list $NewPair$ if it is known to be D3-mergeable but this fact has not yet been utilized for finding new D3-mergeable pairs. We also use the \emph{inverted transition table} $\bfI = (\bfI[a,x])_{a\in A,x\in X}$, where $\bfI[a,x] := \{i\in A \mid a\in \calF(i,x)\}$ for all $a\in A$ and $x\in X$.
The steps of the algorithm are as follows:

1.  Set $M[i,j] := 0$ for $1\leq i<j \leq n$, $NewPair := \ve$ (empty list), and compute $\bfI$.

2. Find all pairs $(a,x)\in A\times X$ for which $|\bfI[a,x]| > 1$. For each such pair $(a,x)$ consider every pair $i,j \in \bfI[a,x]$ with $i<j$. If $M[i,j] = 0$, let $M[i,j] := 1$ and append $(i,j)$ to $NewPair$.

3. Until $NewPair = \ve$ do the following. Delete the first pair, say $(a,b)$, from $NewPair$. For each $x\in X$, find all pairs $(i,j)$ with $i<j$ such that $i \in \bfI[a,x]$ and $j\in \bfI[b,x])$, or $j \in \bfI[a,x]$ and $i\in \bfI[b,x])$. If $M[i,j] = 0$, let $M[i,j] := 1$ and append $(i,j)$ to $NewPair$.

4. $\calF$ is D3-directable if and only if $M[i,j] = 1$ whenever $1\leq i < j \leq n$.

Quite the same way as in \cite{ImSt95}, one can show that the time bound for the algorithm is $\calO(m\cdot n^2)$, where $m = |X|$ and $n= |A|$.

Of course,  the D1- or D2-directability of a fuzzy automaton can be decided by constructing the recognizers described in the proof of Proposition \ref{pr:DW(A) regular}, but this is feasible for very small automata only. It also appears unlikely that anything like the above algorithm  exists for D1- or D2-directability as there does not seem to be any useful notions of D1- or D2-merging words. This may be connected with the fact that, while the shortest D3-directing word of a  complete D3-directable  $n$-state NFA is at most of length of order $\mathcal{O}(n^3)$ (cf. Proposition 5.3 of \cite{ImSt99}), no polynomial upper bounds exist for shortest D1- and D2-directing words (cf. \cite{Bur76,GaIvNG09,GoHeKoRy82,ImSt99}).


\section{Algebraic constructions and directability}\label{se:Alg constr and dir}

As noted in \cite{ImSt95}, subautomata, homomorphic images, and finite direct products of directable DFAs are also directable.
In this section we shall consider the preservation of the various directability properties of fuzzy automata under the corresponding algebraic constructions. The relevant concepts, in different forms, and many of their properties, can be found in \cite{MoMa02}. Some of our formulations follow Petkovi\'{c} \cite{Pet06} who  established for fuzzy automata (with outputs) some fundamental relationships between congruences, homomorphisms and quotient algebras.

In what follows, $\calF$ and $\calG$ always denote the fuzzy automata $(A,X,f)$ and $(B,X,g)$, respectively.

The automaton $\calG$ is a \emph{subautomaton} of $\calF$ if
$B \se A$, $\calF(b,x) \se B$ for all $b\in B$ and $x\in X$, and $g(b,x,b') = f(b,x,b')$ for all $b,b'\in B$ and $x\in X$. Of course, this means that $\calF(b,w) \se B$ and $f(b,w,b') = g(b,w,b')$ for all $b,b'\in B$ and $w\in X^*$. Hence the following lemma.

\begin{lemma}\label{le:F(b,w) = G(b,w)} If $\calG$ is a subautomaton of $\calF$, then $\calF(b,w) = \calG(b,w)$ for all $b\in B$ and $w\in X^*$.
\end{lemma}

\begin{proposition}\label{pr:S(Dir_i) se Dir_i} For any subautomaton $\calG$ of a fuzzy automaton $\calF$ and for each $i = 1,2,3$, $D_i(\calF) \se D_i(\calG)$, and if $\calF$ is $Di$-directable, then so is $\calG$.
\end{proposition}

\bp For each $i =1,2,3$, the second statement follows from the first one, and this we get by Lemma \ref{le:F(b,w) = G(b,w)}. For example, if  $w \in D_3(\calF)$, then there is a state $c\in A$ which appears in every set $\calF(a,w)$ ($a\in A$). In particular, $c\in \calF(b,w) = \calG(b,w)$ for every $b\in B$, and therefore $w\in D_3(\calG)$. \ep

A mapping $\vp : A \ra B$ is a \emph{homomorphism} $\vp : \calF \ra \calG$ from $\calF$ to $\calG$ if \vspace{-3pt}
$$
g(a\vp,x,b) = \max\{f(a,x,a') \mid a'\in A, a'\vp = b\},\vspace{-3pt}
$$
for all $a\in A$, $b\in B$ and $x\in X$. It is an \emph{epimorphism} or an \emph{isomorphism} if it is, respectively, surjective or bijective. If there is an epimorphism $\vp : \calF \ra \calG$, then $\calG$ is called a \emph{homomorphic image} of $\calF$, and if there is an isomorphism from $\calF$ to $\calG$, then $\calF$ and $\calG$ are \emph{isomorphic}, $\calF \cong \calG$ in symbols.

\begin{lemma}\label{le:F(a,w)vp = G(avp,w)} If $\vp : \calF \ra \calG$ is an epimorphism, then $\calF(a,w)\vp = \calG(a\vp,w)$ for all $a\in A$ and $w\in X^*$.
\end{lemma}

\bp If $a'\in \calF(a,w)$, then \vspace{-3pt}
$$
g(a\vp,w,a'\vp) = \max\{f(a,w,a'') \mid a''\vp = a'\vp\} \geq f(a,w,a') > 0 \vspace{-3pt}
$$
implies $a'\vp \in \calG(a\vp,w)$. Hence, $\calF(a,w)\vp \se \calG(a\vp,w)$. Conversely, if $b\in \calG(a\vp,w)$, then
$\max\{f(a,w,a') \mid a'\vp = b\} = g(a\vp,w,b) > 0$ means that
$a'\vp = b$ and $f(a,w,a') > 0$ for some $a'\in A$, and hence $b  \in \calF(a,w)\vp$. \ep

\begin{proposition}\label{pr:H(Dir_i) se Dir_i}
If $\vp : \calF \ra \calG$ is an epimorphism, then for each  $i = 1,2,3$, $D_i(\calF) \se D_i(\calG)$, and if $\calF$ is $Di$-directable, then so is $\calG$.
\end{proposition}

\bp  Again it suffices to prove the first claim for the three values of $i$.

Let $w\in D_1(\calF)$ and assume that $\calF(a,w) = \{c\}$ for every $a\in A$. For every $b\in B$, there is an $a\in A$ with $a\vp = b$, and therefore $\calG(b,w) = \calF(a,w)\vp = \{c\vp\}$, which shows that $w\in D_1(\calG)$.

Let $w\in D_2(\calF)$ and consider any states $b,b'\in B$. If we choose any $a,a'\in A$ satisfying $a\vp = b$ and $a'\vp = b'$, then by Lemma \ref{le:F(a,w)vp = G(avp,w)}, $\calG(b,w) = \calF(a,w)\vp = \calF(a',w)\vp = \calG(b',w)$, which proves that $w\in D_2(\calG)$.

Finally, if $w\in D_3(\calF)$ and $c\in A$ appears in every set $\calF(a,w) \; (a\in A)$, then $a\vp \in \calG(b,w)$ for every $b\in B$, and thus $w\in D_3(\calG)$. \ep

Exactly as for algebras in general (cf. \cite{BuSa81}, for example), the homomorphic images of a fuzzy automaton are, up to isomorphism, its quotient automata. Hence the above results can  be formulated for quotient automata, too.

A \emph{congruence} of a fuzzy automaton $\fzA$ is an equivalence on $A$ such that if $a\,\theta\,a'$  for some $a,a'\in A$, then for all $x\in X$ and $b\in B$,
\vspace{-3pt}
$$
\max\{f(a,x,b') \mid b'\in [b]\} \,=\, \max\{f(a',x,b') \mid b'\in [b]\}.\vspace{-3pt}
$$
It is easy to see that $\theta \in \Eq(A)$ is a congruence of $\calF$ if and only if $A/\theta$ is an \emph{admissible partition} as defined in \cite{MoMa02}, for example.  For a congruence $\theta$, the \emph{quotient automaton} $\calF/\theta = (A/\theta,X,f_\theta)$ is defined by \vspace{-3pt}
$$
f_\theta([a],x,[b]) = \max\{f(a',x,b') \mid a'\in [a], b' \in [b]\} \quad (a,b\in A, x\in X). \vspace{-3pt}
$$
Since $\theta$ is a congruence, $f_\theta([a],x,[b]) = \max\{f(a,x,b') \mid  b' \in [b]\}$.
In \cite{Pet06} Petkovi\'{c} proves that \vspace{-3pt}
\begin{itemize}\vspace{-3pt}
  \item[{\rm (1)}] if $\theta$ is a congruence of $\calF$, then $\nu_\theta : \calF \ra \calF/\theta, a \mapsto [a]$, is an epimorphism, and that \vspace{-3pt}
  \item[{\rm (2)}] if $\vp : \calF \ra \calG$ is an epimorphism, then $\calF/\ker\vp \cong \calG$ for the kernel congruence $\ker\vp := \{(a,a') \mid a,a'\in A, a\vp = a'\vp\}$ of $\calF$. \vspace{-3pt}
\end{itemize}
\vspace{-3pt}
These facts yield the following corollary of Proposition \ref{pr:H(Dir_i) se Dir_i}.

\begin{corollary}\label{co:Quotient(Dir_i) se Dir_i} For any congruence $\theta$ of a fuzzy automaton $\calF$ and each  $i = 1,2,3$, $D_i(\calF) \se D_i(\calF/\theta)$, and if $\calF$ is $Di$-directable, then so is $\calF/\theta$.
\end{corollary}

Since any direct product of finitely many factors can be obtained, up to isomorphism, by forming products of two factors, we may restrict ourselves to products of two factors.
The \emph{direct product} of $\calF$ and $\calG$ is the fuzzy automaton $\calF \times \calG = (A \times B,X,h)$, where $h$ is defined by \vspace{-3pt}
$$
h((a,b),x,(a',b')) = \min\{f(a,x,a'),g(b,x,b')\} \quad (a,a'\in A, b,b'\in B, x\in X). \vspace{-3pt}
$$

\begin{lemma}\label{le:Direct products} For any $A'\se A$, $B'\se B$, $x\in X$, $a\in A$, $b\in B$ and $w\in X^*$,
\begin{itemize}\vspace{-3pt}
  \item[{\rm (a)}] $(\calF\times\calG)(A'\times B',x) = \calF(A',x) \times \calG(B',x)$, and \vspace{-3pt}
  \item[{\rm (b)}] $(\calF\times\calG)((a,b),w) = \calF(a,w) \times \calG(b,w)$.
\end{itemize}
\end{lemma}

\bp For any $a'\in A$, $b'\in B$ and $x\in X$, \vspace{-3pt}
\begin{align*}
       (a',b')\in (\calF\times\calG)(A'\times B',x)\; &\mathrm{iff} \;(\exists (a,b)\in A'\times B')f(a,x,a'), g(b,x,b') > 0\\
       &\mathrm{iff} \; (\exists a\in A')a'\in \calF(a,x), (\exists b\in B')b'\in \calG(b,x)\\
        &\mathrm{iff} \; (a',b') \in \calF(A',x)\times \calG(B',x),\vspace{-3pt}
\end{align*}
which proves (a).
Statement (b) can now be proved by induction on $\lg(w)$. The case $w = \ve$ is trivial, and for $w\in X$, the equality follows directly from the definition of $h$. Let $w = ux$ with $u\in X^*$ and $x\in X$, and assume that $(\calF\times\calG)((a',b'),u) = \calF(a',u)\times\calG(b',u)$ for all $a'\in A', b'\in B'$. Then \vspace{-3pt}
\begin{align*}
       (\calF\times\calG)((a,b),w)\; &= \; (\calF\times\calG)((\calF\times\calG)((a,b),u),x)\\
       &= \; (\calF\times\calG)(\calF(a,u)\times \calG(b,u),x)\\
       &= \; \calF(\calF(a,u),x)\times \calG(\calG(b,u),x)\\
       &= \; \calF(a,w)\times \calG(b,w), \vspace{-3pt}
\end{align*}
for all $(a,b)\in A\times B$, where we also used (a). \ep

The following example shows that D1-directability is not preserved by direct products even for complete fuzzy automata.

\begin{example}\label{ex:D1 and direct product}{\rm Let $\calF = (\{a,b\},\{x,y\},f)$ and $\calG = (\{1,2\},\{x,y\},g)$ be any complete fuzzy automata with $\calF(a,x) = \calF(b,x) = \{b\}$, $\calF(a,y) = \{a\}$, $\calF(b,y) = \{a,b\}$, $\calG(1,x) = \{1\}$, $\calG(2,x) = \{1,2\}$, and  $\calG(1,y) = \calG(2,y) = \{2\}$. Then $x\in D_1(\calF)$ and $y\in D_1(\calG)$, but $\calF\times\calG$ has no D1-directing word since $\calF(b,uy) = \{a,b\}$ and $\calG(2,ux) = \{1,2\}$ for every $u\in X^*$.}
\end{example}

For D2-directability, the following holds.

\begin{proposition}\label{pr:D2-dir and direct product} The direct product of any two complete D2-directable fuzzy automata is also D2-directable.
\end{proposition}

\bp Let $u\in D_2(\calF)$ and $v\in D_2(\calG)$, and let $C = \calF(a,u)$ for every $a\in A$ and $D = \calG(b,v)$ for every $b\in B$. By Lemma \ref{le:Direct products}, $(\calF\times\calG)((a,b),uv) =  \calF(C,v)\times D$ for every $(a,b)\in A\times B$, and hence $uv \in D_2(\calF\times\calG)$. \ep

Proposition \ref{pr:D2-dir and direct product} does not hold without the completeness assumption.

\begin{example}\label{ex:D2 and incompl}{\rm Let $\calF = (\{a,b\},\{x,y\},f)$ and $\calG = (\{1,2\},\{x,y\},g)$ be any fuzzy automata for which $\calF(a,x) = \calF(b,x) = \calF(b,y) = \{b\}$, $\calF(a,y) = \es$,  $\calG(1,y) = \calG(2,y) =  \calG(2,x) = \{2\}$, and  $\calG(1,x) = \es$. Then $x\in D_2(\calF)$ and $y\in D_2(\calG)$, but $\calF\times\calG$ has no D2-directing word because $(\calF\times\calG)((a,1),yu) = (\calF\times\calG)((a,1),xu) = \es$ while $(\calF\times\calG)((b,2),yu) = (\calF\times\calG)((b,2),xu) = \{(b,2)\}$ for every $u\in X^*$.}
\end{example}

\begin{proposition}\label{pr:D3 and direct product} The direct product of any two complete D3-directable fuzzy automata is also D3-directable.
\end{proposition}

\bp Let $u\in D_3(\calF)$ and $v\in D_3(\calG)$, and let $a_0\in A$ and $b_0\in B$ be states such that $a_0\in \calF(a,u)$ for every $a\in A$ and $b_0\in \calG(b,v)$ for every $b\in B$. Pick any $a_1$ from $\calF(a_0,v)$; this is possible as $\calF$ is complete. Then \vspace{-3pt}
$$
(a_1,b_0) \in \calF(a,uv)\times\calG(b,uv) = (\calF\times\calG)((a,b),uv) \vspace{-3pt}
$$
for any $(a,b)\in A\times B$, which shows that $uv\in D_3(\calF\times\calG)$. \ep

That Proposition \ref{pr:D3 and direct product} does not hold without the completeness assumption, can be seen considering the automata $\calF$ and $\calG$ of Example \ref{ex:D2 and incompl}: they are both D3-directable, but $\calF\times\calG$ is not.

It is easy to see that subautomata, homomorphic images and finite direct products of complete fuzzy automata are complete. We may therefore sum up the positive results of this section as follows.

\begin{proposition}\label{pr:Closure prop of FDir(i) and CFDir(i)} For every $i = 1,2,3$, the classes $\FDir(i)$ and $\CFDir(i)$ are closed under subautomata and homomorphic images, and the classes $\CFDir(2)$ and $\CFDir(3)$ are closed also under finite direct products.
\end{proposition}

\section{Concluding remarks: possible extensions}\label{se:Concluding}
We have considered classical max-min fuzzy automata over the real interval $[0,1]$. Moreover, all the directing words considered were ``location-synchronizing'' in the sense of \cite{DoJuLaMaSh13} in that just the states reachable by a word are taken into account while the transition weights are ignored. Therefore many results could be deduced from the corresponding facts about usual nondeterministic finite automata. Obviously, there are weighting spaces and alternative notions of directability for which such a reduction to nondeterministic automata is not useful anymore, but let us consider some possible extensions of the theory in which NFAs still could be used.

A natural generalization  of Zadeh's \cite{Zad65} original notion of fuzziness is obtained by replacing the interval $[0,1]$ by a general bounded distributive lattice $L$. In our setting an $L$\emph{-fuzzy automaton} is a system $\fzA$, where $A$ and $X$ are as before, but the transition relation is a mapping $f : A\times X \times A \ra L$. Such automata (with initial and final distributions) are also called \emph{multivalued automata} (cf. \cite{Rah09}). Let $\vee$ and $\wedge$ denote the join- and meet-operators of $L$, and let $0$ be the least and $1$ be the greatest element of $L$. The extension $f^*: A\times X^* \times A \ra L$ of $f$ is then defined as follows:

(1) $f^*(a,\ve,a) = 1$, and $f^*(a,\ve,b) = 0$ if $b\neq a$, for all $a,b\in A$, and

(2) $f^*(a,w,b) = \bigvee\{f^*(a,v,c)\wedge f(c,x,b) \mid c\in A\}$ for all $a,b\in A$ and $w = vx$ with $v\in X^*$ and $x\in X$.

Furthermore, let $\calF(a,w) := \{b\in A \mid f^*(a,w,b) > 0\}$ for all $a\in A$ and $w\in X^*$. The defining conditions (D1'), (D2') and (D3') of D1-, D2- and D3-directing words can now be applied as such to any $L$-fuzzy automaton $\calF$. If $L$ is a chain, then $p\vee q = \max(p,q)$ and $p\wedge q = \min(p,q)$ for all $p,q \in L$, and exactly the same way as for our basic fuzzy automata we may associate with an $L$-fuzzy automaton $\calF$ an NFA $\hF$ which has the same D$i$-directing words ($i=1,2,3$) as $\calF$. More generally, this can be done whenever $L$ is a lattice in which $p\wedge q = 0$ only in case $p=0$ or $q=0$.

To require transition weights exceeding some \emph{threshold} $\tau,\, 0\leq \tau < 1$ is a simple way to add a quantitative component to our notions of directability. For any fuzzy automaton $\fzA$, and any $w\in X^*$ and $a\in A$, let $\calF_\tau(a,w) := \{b\in A \mid f(a,w,b)>\tau\}$. Next we modify conditions (D1'), (D2') and (D3') as follows: a word $w\in X^*$ is said to be
\begin{itemize} \vspace{-3pt}
  \item[{\rm (1)}] \emph{D1,$\tau$-directing} if $(\exists c\in A)(\forall a\in A)(\calF_\tau(a,w) = \{c\} )$, \vspace{-3pt}
  \item[{\rm (2)}] \emph{D2,$\tau$-directing} if $(\forall a,b\in A)(\calF_\tau(a,w) = \calF_\tau(b,w))$, and\vspace{-3pt}
  \item[{\rm (3)}] \emph{D3,$\tau$-directing} if $(\exists c\in A)(\forall a\in A)(c\in \calF_\tau(a,w))$. \vspace{-3pt}
\end{itemize}
It is easy to see that for each $i=1,2,3$, the D$i,\!\tau$-directing  words of $\calF$ are the same as the D$i$-directing words of the NFA $\calF_\tau = (A,X)$ defined by $ax^{\calF_\tau} := \calF_\tau(a,x)$ ($a\in A$, $x\in X$). Hence, the theory of directable NFAs is again applicable.

\end{document}